\documentclass[prl,twocolumn,superscriptaddress]{revtex4-2}
\usepackage{amsmath,amssymb,amsfonts,mathrsfs} 	% Typical maths resource package
\usepackage{graphicx}
\usepackage{subfigure}
\usepackage{xcolor}
\usepackage[normalem]{ulem}
\renewcommand{\[}{\begin{equation}}
\renewcommand{\]}{\end{equation}}
\def\bea{\begin{eqnarray}}
\def\eea{\end{eqnarray}}
\def\nn{\nonumber\\}

\newcommand{\emi}[1]{{\rm e}^{-i #1}}
\newcommand{\ei}[1]{{\rm e}^{i #1}}

\newcommand{\p}{{\bf p}}

\renewcommand{\k}{{\bf k}}

\renewcommand{\r}{{\bf r}}

\newcommand{\equ}[1]{Eq.~(\ref{#1})}
\newcommand{\eqs}[2]{Eqs.~(\ref{#1}) and (\ref{#2})}

\def\ket#1{\vert#1\rangle}
\def\ev#1{\langle#1\rangle}
\def\me#1#2#3{\langle#1| \, #2 \, |#3\rangle}
\def\runtime{(\the\time)\qquad\the\month/\the\day/\the\year}% get current time
\def\today
 {\count10=\year\advance\count10 by -2000 \number\day--\ifcase
  \month \or Jan\or Feb\or Mar\or Apr\or May\or Jun\or
             Jul\or Aug\or Sep\or Oct\or Nov\or Dec\fi--\number\count10}

\def\hour{\count10=\time\count11=\count10
\divide\count10 by 60 \count12=\count10
\multiply\count12 by 60 \advance\count11 by -\count12\count12=0
\number\count10 :\ifnum\count11 < 10 \number\count12\fi\number\count11}

\begin{document}

%%%%%%%%%%%%%%%%%%%%%%%%%%%%%%%%%%%%%%%%%
\title{Geometrical theory of the shift current in presence of disorder and interaction}

\author{Raffaele Resta}
\email{resta@iom.cnr.it.it}
\affiliation{CNR-IOM Istituto Officina dei Materiali, Strada Costiera 11, 34149 Trieste, Italy}
\affiliation{Donostia International Physics Center, 20018 San Sebasti{\'a}n, Spain}

\date{\today}

\begin{abstract} The electric field of light induces---in a non centrosymmetric insulator---a dc current, quadratic in the field magnitude, and called ``shift current''. When addressed from a many-electron viewpoint, the shift current has a simple explanation and a simple formulation as well, deeply rooted in quantum geometry. The basic formula is then specialized to the independent-electron case, first for a disordered system in a supercell formulation, and then for a crystalline system. In the latter case the known shift-current formula is retrieved in a very transparent way.
\end{abstract}

\date{run through \LaTeX\ on \today\ at \hour}

%\pacs{xxx}

\maketitle %\bigskip\bigskip

In materials lacking inversion symmetry an electric field at frequency $\omega$ generates a dc current, quadratic in the field magnitude: the phenomenon goes under the name of ``shift current''; its theory is formulated  for a crystalline system of independent electrons, and is rooted in the quantum geometry of Bloch electrons \cite{Belinicher82,Sipe00,Young12,Morimoto16,Fregoso17,Cook17,Ibanez18,Dai23}. The phenomenon has also been observed in presence of strong disorder \cite{Hatada20}. Its robustness has been regarded as a manifestation of the geometrical nature of the shift current beyond band-structure theory  \cite{Nagaosa22}, but an actual geometrical formulation for the general case is  lacking so far. Such a theory is provided here; its main entry is the many-body shift vector, a simple geometrical entity closely related to macroscopic polarization. The present general theory copes with interacting electrons and noncrystalline materials; in the crystalline case the known formula is retrieved in a very transparent way.

The shift current has a perspicuous explanation when the solid as a whole is regarded from a many-electron viewpoint: the light pumps the population of the resonant excited states at a constant rate, proportional to $E^2(\omega)$. 
The excited states have a different polarization from the ground state; the induced currents---i.e. the time-derivative of the polarizations---are therefore equal to the rate times the polarization differences; as shown in this Letter, the latter have an elegant quantum-geometrical expression. 
The shift current is a second-order response, nonetheless---when addressed in this way---its explanation does not require second-order perturbation theory, at variance with the existing derivations in the literature. The present  picture applies to linearly polarized light only, while circularly polarized light and ``injection current'' are not addressed; see however a comment in the conclusions of this Letter.

In order to simplify the presentation, I assume a material where inversion symmetry is absent, but which is symmetric by reflection in the $x$ and $y$ directions; I only address the most paradigmatic component of the quadratic conductivity, which for linearly polarized light is $\sigma_{zzz}^{(2)}(0;\omega,-\omega)$. Therefore
all the relevant vectors (field, polarization, momenta, current\dots) are directed along $z$ and scalar-like notations can be safely adopted; the shorthand notation \[ \tilde\sigma(\omega) = \sigma_{zzz}^{(2)}(0;\omega,-\omega) \] is also adopted throughout. The linearly polarized time-dependent field is $E(t) = E(\omega) ( \emi{\omega t} +  \emi{\omega t} )$ and the shift current is \cite{Fregoso17,Ibanez18}: \[ j = 2 \; \tilde\sigma(\omega) E^2(\omega) . \label{s1} \]

One considers a system of $N$ interacting electrons in a cubic box of volume $L^3$, whose eigenstates obey Born-von-K\`arm\`an periodic boundary conditions (PBCs). It is expedient to adopt the family of many-body Hamiltonians parametrized by $\kappa$: \[ \hat{H}_{\kappa} = \frac{1}{2m} \sum_{i=1}^N \left(\p_i + \hbar \kappa {\bf e}_z \right)^2 + \hat{V}, \label{kohn} \] where $\hat{V}$ includes the one-body potential and electron-electron interaction, and ${\bf e}_z$ is the unit vector in the $z$-direction. The $\kappa$-dependent eigenstates are normalized in the hypercube of volume $L^{3N}$. 
The limit $N \rightarrow \infty$, $L \rightarrow \infty$, $N/L^3$ constant is understood. The $\kappa$-dependent term in the kinetic energy is generally dubbed ``flux'' or ``twist'': it is the main parameter entering geometrical and topological properties in a many-body framework, which includes cases with interaction and disorder \cite{Ortiz94,Xiao10}.

According to the theory of semiclassical absorption, the  population $\pi_n$ of the $n$-th excited state  increases linearly in time in presence of an electric field at the resonance frequency; the ground state is depopulated by the same amount. If $P_0$ and $P_n$ are the macroscopic polarizations of the many-body states $\ket{\Psi_0}$ and $\ket{\Psi_n}$, respectively, then the induced polarization is \[ P^{(\rm ind)}(t) = \sum_{n \neq 0} \pi_n(t) \Delta P_{0n} \qquad  \Delta P_{0n} = P_n - P_0 . \label{d1} \] 

Elementary time-dependent perturbation theory yields \cite{Ballentine} \[ \pi_n(t) = \frac{2\pi e^2}{\hbar^2}E^2(\omega)| \me{\Psi_0}{\hat{z}}{\Psi_n}|^2 \delta(\omega -\omega_{0n}) \, t , \label{balle} \] where $\omega_{0n} = (E_n - E_0)/\hbar$. Such expression goes under the name of Fermi's golden rule;
here  the dipole matrix elements---within PBCs---are defined as \[  \me{\Psi_0}{\hat{z}}{\Psi_n} = i \frac{ \me{\Psi_0}{\sum_i p_i}{\Psi_n}}{m\omega_{0n}}. \label{comm} \] 
 Given that $j = \dot{P}^{(\rm ind)}$, a steady current flows therefore through the sample; from Eqs. (\ref{s1}), (\ref{d1}), and (\ref{balle}) one gets 
 \[ \tilde\sigma(\omega) =   \frac{\pi e^2}{\hbar^2} \sum_{n  \neq 0} | \me{\Psi_0}{\hat{z}}{\Psi_n}|^2 \Delta P_{0n} \delta(\omega -\omega_{0n}) . \label{s2} \] 
As usual when dealing with conductivities, the singular function becomes a continuous function in the $L \rightarrow \infty$ limit \cite{rap165}.

Next, an expression for $\Delta P_{0n}$ is needed. It is known since the dawn of polarization theory in the early 1990s  \cite{rap73,King93} that dealing with polarization differences is easier than dealing with polarization itself;  this proves to be the case even here. 

We abandon for a while PBCs, and we assume that $\ket{\Psi_n}$ and $\ket{\Psi_0}$ are instead eigenstates of \equ{kohn} for a bounded crystallite within open boundary conditions (OBCs): then obviously \[ \Delta P_{0n} = -\frac{e}{L^3} (\me{\Psi_n}{\hat{z}}{\Psi_n} - \me{\Psi_0}{\hat{z}}{\Psi_0}),  \label{forbid} \] where 
  $\hat{z} = \sum_i z_i$ . This expression does not make sense within PBCs, because the diagonal elements of the operator $\hat{z}$ are ill defined  therein \cite{rap100}; after all, this is the reason why the dipole of a crystallite and the polarization of a solid look so different theory-wise. 
    
Therefore the strategy---inspired by Ref. \cite{rap162}---is to find an alternative expression for  $\Delta P_{0n}$ which makes sense within both OBCs and PBCs and can therefore be adopted on the same ground in both cases. To this aim we notice that the flux in \equ{kohn} is a vector potential, constant in space and time, i.e. a pure gauge; we also remind that PBCs violate gauge-invariance \cite{Kohn64}, while instead OBCs do not. In the latter case the gauge-transformed state vectors are  \[ \ket{\Psi_{n\kappa}} = \ei{\phi_n(\kappa)}\emi{\kappa \hat{z}} \ket{\Psi_{n}} ,\]  and the polarization $P_n$ of the $n$-th state can  be identically rewritten, at any $\kappa$, in terms of a Berry connection:  \[ P_n =  -\frac{ie}{L^3}\ev{\Psi_{n\kappa} | \partial_\kappa \Psi_{n\kappa}} -\frac{ie}{L^3} \phi'_n(\kappa) . \] This expression cannot be adopted as such in evaluating $\Delta P_{0n}$, due to the arbitrary gauge phases $\phi_n(\kappa)$ and $\phi_0(\kappa)$. The drawback is removed by defining the many-body shift vector as \bea {\cal R}_{0n} &=&  i \ev{\Psi_n|\partial_\kappa \Psi_n} -  i \ev{\Psi_0 |\partial_\kappa \Psi_0} \nn &+& \partial_\kappa \mbox{Im ln } \ev{\Psi_0 | \partial_\kappa \Psi_n} \label{diff} , \eea  where $\kappa$ has been set to zero for the sake of simplicity. \equ{diff} is a manifestly gauge-invariant geometrical entity,
where the second line cancels the arbitrary gauge phases. One thus arrives at
 \[ \Delta P_{0n} = -\frac{ie}{L^3} {\cal R}_{0n}\label{this} ; \]
this expression, derived within OBCs, is adopted as such within PBCs.

Notice that I am adopting here the same logic as  for \equ{comm}, which indeed in the OBCs case is an identity, while in the PBCs case  is a {\it definition} of the off-diagonal elements of the  $\hat{z}$ operator. Quite analogously it has been proved here that \equ{this} is an identity within OBCs; it becomes a definition within PBCs. The ``forbidden'' position operator appearing in \equ{forbid}  is here tamed by adopting in \equ{this} exactly the same logic as routinely done for \equ{comm}. 

 By inserting \equ{this} in \equ{s2} we finally obtain the compact expression: \[ \tilde\sigma(\omega) =   - \frac{\pi e^3}{\hbar^2 L^3} \sum_{n  \neq 0} | \me{\Psi_0}{\hat{z}}{\Psi_n}|^2 {\cal R}_{0n} \, \delta(\omega -\omega_{0n}) , \label{s3} \] which is the major result of the present Letter. The $\ket{\Psi_n}$ are the most general excited states of the $N$-electron insulating system;
 the virtue of  \equ{s3} is of being very general and formally exact, while the excited states $\ket{\Psi_n}$ of an explicitly correlated Hamiltonian are notoriously challenging on the computational side. 
 
 Most of the shift-current literature is rooted in band-structure theory for noninteracting electrons. A few papers considered  renormalizing the single-particle energies to include interaction effects \cite{Chaudhary22} and  electron-hole contributions to both dipole matrix elements and  excitation energies \cite{Pedersen15,Fei20}; the latter approach also allows for excitonic subbandgap shift currents \cite{Chan21}. The present quite general formulation includes in principle such kind of excitations; furthermore, given its simplicity, it might provide the basis for the future development of some other approximations.
 
In a mean-field  framework (either Hartree-Fock or Kohn-Sham) both the ground state and the excited states have a simple form; \equ{s3} assumes a simple form as well, 
 for both  disordered  and crystalline systems. The remaining of this Letter is devoted to these cases.

 The ground state $\ket{\Psi_0}$ of a system of independent electrons is a Slater determinant of $N/2$ doubly-occupied periodic orbitals $\ket{u_j}$ with energies $\epsilon_j$; the generic excited state $\ket{\Psi_n}$ is a monoexcited determinant where the occupied state $j$ is replaced by the unoccupied $j'$ one; the transition frequency $\omega_{0n}$ becomes then $\omega_{jj'} = (\epsilon_{j'} - \epsilon_j)/\hbar$. One thus recasts \equ{d1} as \[ P^{(\rm ind)}(t) =   \!\!\!\!\!\!\! \sum_{\substack{j={\rm occupied } \\ j'={\rm unoccupied}} } \!\!\!\!\!\! \pi_{jj'}(t) \Delta P_{jj'} ; \]  \[  \pi_{jj'}(t) =  \frac{2\pi e^2}{\hbar^2}E^2(\omega)| \me{u_j}{z}{u_{j'}} |^2 \delta(\omega -\omega_{jj'}) \, t . \label{balles} \] 
The many-body matrix element in \equ{balle} has been transformed---owing to the Slater-Condon rules---into the matrix element of the corresponding one-body operator, i.e. \[ \me{u_j}{z}{u_{j'}} = i \frac{\me{u_j}{p}{u_{j'}} }{m \omega_{jj'}} . \]  The Berry connection of a Slater determinant equals the sum of the Berry connections of the orbitals \cite{rap_a20}, ergo (for spinless electrons) \[ i \ev{\Psi_n|\partial_\kappa \Psi_n} -  i \ev{\Psi_0 |\partial_\kappa \Psi_0 } =  i \ev{u_{j'}|\partial_\kappa u_{j'}} -  i \ev{u_j |\partial_\kappa u_j} .
\] The shift vector becomes then \[ R_{jj'} =  i \ev{u_{j'}|\partial_\kappa u_{j'}} -  i \ev{u_j |\partial_\kappa u_j} +  \partial_\kappa \mbox{Im ln } \ev{u_j | \partial_\kappa u_{j'}} .
\label{ss4} \] From the above results the independent-particle version of \equ{s3} is \[ \tilde\sigma(\omega) = - \frac{\pi e^3}{\hbar^2 L^3} \!\!\!\!\!\!\! \sum_{\substack{j={\rm occupied } \\ j'={\rm unoccupied}} } \!\!\!\!\!\! |  \me{u_j}{z}{u_{j'}} |^2 R_{jj'} \delta(\omega -\omega_{jj'}) . \label{s4}  \]  This formula could be implemented as such within DFT in a supercell framework, to deal with a disordered (yet non inversion-symmetric) material, such as a solid solution with chemical disorder. To the best of the Author's knowledge, the disorder issue has only been investigated via model Hamiltonians \cite{Ishizuka21,Nagaosa22}.

%Any differentiable gauge can be adopted, including the parallel-transport gauge, in which case only the third term in the shift vector yields a nonvanishing contribution.

I switch from now on to a crystalline insulator. After the $L \rightarrow \infty$ limit is taken the Bloch vector $\k$ becomes continuous; the spectrum becomes continuous as well. The single-particle orbitals are Bloch states $\ket{\psi_{j\k}} = \ei{\k \cdot \r} \ket{u_{j\k}}$, normalized to one in the crystal cell. I start addressing the simple case where only two bands contribute to the shift current.
The relevant band energies are $\hbar \omega_{0\k}$ and $\hbar \omega_{1\k}$; in the ground state the lowest band is occupied and the highest band is empty.

Let us preliminarily consider the simple case where all the electrons in the lowest band are promoted to the highest one. Polarization theory yields (for spinless electrons) the polarization difference 
\bea \Delta P &=& i e  \int_{\rm BZ} \frac{d \k}{(2\pi)^3} \ev{u_{0\k} |\partial_{k_z}  u_{0\k} } \nn &-&  i e \int_{\rm BZ} \frac{d \k}{(2\pi)^3} \ev{u_{1\k} |\partial_{k_z}  u_{1\k} }  , \label{ksv} \eea where BZ is the Brillouin zone. Being the difference of two Berry phases, \equ{ksv} is affected by the notorious ``quantum'' indeterminacy  in each of them \cite{Vanderbilt}. The drawback is removed by making use of the band-structure shift vector \bea R_{01}(\k) &=&  i\ev{u_{1\k} |\partial_{k_z}  u_{1\k} } - i\ev{u_{0\k} |\partial_{k_z}  u_{0\k}} \nn &+& \partial_{\k} \mbox{Im ln }  \ev{u_{0\k} |\partial_{k_z}  u_{1\k}} , \label{sk} \eea whose second line integrates to zero over the BZ  (the periodic gauge is assumed \cite{Vanderbilt}). The polarization difference is then rewritten as \[ \Delta P = -e \int_{\rm BZ} \frac{d \k}{(2\pi)^3} R_{01}(\k) ,\label{rapix} \] and \eqs{ksv}{rapix} are {\it apparently} equivalent. There is an outstanding difference, though: the integrands in \equ{ksv} are gauge-dependent, while the one in \equ{rapix} is gauge-invariant, and $\Delta P$ is no longer  affected by the quantum indeterminacy. The extra term in  \equ{sk} only fixes the gauge in \equ{ksv}, while adding no extra contribution.
It may happen that the matrix element in the second line of \equ{sk}  vanishes at some points in the BZ, thus making its phase ill defined \cite{Fregoso17}; arguably, this would only happen in a zero-measure set, not spoiling the argument in the three-dimensional case.

The fact that $\Delta P$ in \equ{rapix} is the BZ-integral of a gauge-invariant integrand is of overwhelming importance: $\Delta P$ becomes the $\k$-resolved sum of individual contributions originating  from vertical interband transitions. The radiation selectively pumps only the resonant ones  and therefore: \[ P^{(\rm ind)}(t) = -e \int_{\rm BZ} \frac{d \k}{(2\pi)^3} \pi_{01}(\k,t) R_{01}(\k) ; \label{rapix2} \]
 \[ \pi_{01}(\k,t) = \frac{2\pi e^2}{\hbar^2}E^2(\omega)  | \me{\psi_{0\k}}{z}{\psi_{1\k} }|^2  \delta(\omega -\omega_{01\k}) \, t , \label{balle2} \] where
 $\omega_{01\k} = \omega_{1\k} - \omega_{0\k}$.
 %Here again the matrix element has to be understood as \[ \me{\psi_{0\k}}{z}{\psi_{1\k} } = i \frac{\me{\psi_{0\k}}{p}{\psi_{1\k} }}{m(\omega_{1\k} - \omega_{0\k})} . \] 
 As above, by relating $\tilde\sigma(\omega)$ to $\dot{P}^{(\rm ind)}$ one retrieves in a very transparent way the standard shift-current formula for the simple two-band case.  
 %\begin{widetext}\[ \tilde\sigma(\omega) = - \frac{\pi e^3}{\hbar^2} \int_{\rm BZ} \frac{d \k}{(2\pi)^3}  | \me{\psi_{0\k}}{z}{\psi_{1\k} }|^2   R_{01}(\k) \delta(\omega -\omega_{1\k} + \omega_{0\k}) \label{final} .\] \end{widetext} 
 By considering as constant the matrix element therein one retrieves the approximation proposed in Ref. \onlinecite{Fregoso17}.
 
 In a semiclassical picture of the shift current, interband population coherence  plays a relevant role \cite{Fregoso19}. Coherence is in fact implicit in \eqs{balle}{balle2}, given the conditions under which the formula is derived \cite{Ballentine}, namely $t \gg 1/ \omega$. Optical pumping is indeed a coherent phenomenon; in the present case $t$ is the macroscopic time in which the charge transported by the shift current is harvested.
 
 \equ{rapix2} is easily generalized beyond the two-band model by summing the contributions of any pairs of bands whose states are resonating at frequency $\omega$. If the band energies are $\epsilon_{j\k}$ and $\omega_{jj'\k} = (\epsilon_{j'\k} - \epsilon_{j\k})/\hbar$, one gets the known shift-current formula, i.e. 
 \begin{widetext}
 \[ \tilde\sigma(\omega) = - \frac{\pi e^3}{\hbar^2}  \!\!\!\!\!\!\! \sum_{\substack{j={\rm occupied } \\ j'={\rm unoccupied}} } \!\!\!  \int_{\rm BZ} \frac{d \k}{(2\pi)^3}  | \me{\psi_{j\k}}{z}{\psi_{j'\k} }|^2   R_{jj'}(\k) \delta(\omega -\omega_{jj'\k}) \label{final} ;\] \end{widetext} I stress that  its previously known derivations invariably require a considerable amount of algebra.
 
 For practical computations the integral can be discretized in the standard way, since the integrand is gauge invariant; any differentiable gauge can be adopted. This is at variance with the Berry-phase integrals of polarization theory, which instead require a peculiar discretization \cite{Vanderbilt}. If---for a large supercell---\equ{final} is discretized with the $\Gamma$ point only, one retrieves \equ{s4}.
 
 In conclusion, I have addressed the shift current beyond the customary band-structure framework \cite{Belinicher82,Sipe00,Young12,Morimoto16,Fregoso17,Cook17,Ibanez18,Dai23} and its extensions  \cite{Pedersen15,Fei20,Chan21}, where the formulas were obtained by manipulating the cumbersome expansions of second-order perturbation theory. As shown here, both in the many-body case and in the band-structure case the shift current admits a simple expression in terms of Fermi's golden rule. It is worth observing that another second order response, the injection current (not addressed here), is known to  admit a formulation in terms of Fermi's golden rule \cite{Fregoso19}. The main entry in the present many-body formulation is geometrical: the many-body shift vector. It is closely related to polarization theory in its many-body flavor \cite{Ortiz94,rap100,rap162}, and has a perspicuous physical meaning. It measures in fact the difference between the (extensive) dipoles of an excited state and of the ground state.
The independent-electron expression for the shift current, \eqs{ss4}{s4}, is implementable as such to deal with disordered solids (where inversion symmetry is broken) in a supercell framework, similarly in spirit to the ab-initio calculations of infrared spectra \cite{Tuckerman02}. Last but not least, the present theory---when specialized to a crystalline system of independent electrons---leads to the well known formula in a few lines.

I thank Ivo Souza for the many fruitful conversations we had about this topic.
Work supported by the Office of Naval Research (USA) Grant No. N00014-20-1-2847.

%\bibliography{rapix.bib}


\end{document}